\documentclass[lettersize,journal,11pt]{IEEEtran}

\usepackage{enumitem}
\usepackage{svg}
\usepackage{epstopdf}
\usepackage{hyperref}
\usepackage{booktabs}
\usepackage{amsmath,amsfonts}
\usepackage{algorithm}
\usepackage{algpseudocode}
\usepackage{array}
\usepackage[caption=false,font=normalsize,labelfont=sf,textfont=sf]{subfig}
\usepackage{textcomp}
\usepackage{stfloats}
\usepackage{url}
\usepackage{verbatim}
\usepackage{graphicx}
\usepackage{cite}
\usepackage[final]{pdfpages}
\usepackage{multirow}
\usepackage{threeparttable}
\usepackage{marvosym}
\usepackage{tabularx}
\usepackage{soul}
\usepackage{xspace}
\usepackage{tikz}
\usepackage{pifont}
\usepackage{makecell}
\usepackage{setspace}
\usepackage{multicol}
\usepackage{pgfplots}
\usepackage{calligra}
\usepackage{algpseudocode}
\usepackage{caption}
\usepackage{CJK}
\usepackage[numbers]{natbib}

\newcommand*{\circled}[1]{\lower.7ex\hbox{\tikz\draw (0pt, 0pt) circle (.5em) node {\makebox[1em][c]{\small #1}};}}

\DeclareUnicodeCharacter{2642}{\male}
\DeclareUnicodeCharacter{2640}{\female}
\DeclareUnicodeCharacter{2766}{\ding{166}}
\DeclareUnicodeCharacter{2665}{\ding{170}}

\errorcontextlines\maxdimen
\makeatletter
\newcommand*{\algrule}[1][\algorithmicindent]{\makebox[#1][l]{\hspace*{.5em}\thealgruleextra\vrule height \thealgruleheight depth \thealgruledepth}}%
\newcommand*{\thealgruleextra}{}
\newcommand*{\thealgruleheight}{.75\baselineskip}
\newcommand*{\thealgruledepth}{.25\baselineskip}
\newcount\ALG@printindent@tempcnta
\def\ALG@printindent{%
    \ifnum \theALG@nested>0
        \ifx\ALG@text\ALG@x@notext
        \else
            \unskip
            \addvspace{-1pt}
            \ALG@printindent@tempcnta=1
            \loop
                \algrule[\csname ALG@ind@\the\ALG@printindent@tempcnta\endcsname]%
                \advance \ALG@printindent@tempcnta 1
                \ifnum \ALG@printindent@tempcnta<\numexpr\theALG@nested+1\relax
            \repeat
        \fi
    \fi
}
\patchcmd{\ALG@doentity}{\noindent\hskip\ALG@tlm}{\ALG@printindent}{}{\errmessage{failed to patch}}
\makeatother
\newbox\statebox
\newcommand{\myState}[1]{%
    \setbox\statebox=\vbox{#1}%
    \edef\thealgruleheight{\dimexpr \the\ht\statebox+1pt\relax}%
    \edef\thealgruledepth{\dimexpr \the\dp\statebox+1pt\relax}%
    \ifdim\thealgruleheight<.75\baselineskip
        \def\thealgruleheight{\dimexpr .75\baselineskip+1pt\relax}%
    \fi
    \ifdim\thealgruledepth<.25\baselineskip
        \def\thealgruledepth{\dimexpr .25\baselineskip+1pt\relax}%
    \fi
    \State #1%
    \def\thealgruleheight{\dimexpr .75\baselineskip+1pt\relax}%
    \def\thealgruledepth{\dimexpr .25\baselineskip+1pt\relax}%
}

\usepackage{setspace}
\begin{document}

\title{Fed-{\scshape urlBERT}: Client-side Lightweight Federated Transformers for URL Threat Analysis}
\author{Yujie Li,~\IEEEmembership{}
Yanbin Wang\textsuperscript{\Letter},~\IEEEmembership{Member,~IEEE,}
        Haitao Xu\textsuperscript{\Letter},~\IEEEmembership{}
        Zhenhao Guo,~\IEEEmembership{}
        Fan Zhang,~\IEEEmembership{}
        Ruitong Liu,~\IEEEmembership{}
        Wenrui Ma~\IEEEmembership{}

\thanks{Y. Li, Y. Wang, H. Xu, Z. Guo and F. Zhang are affiliated with the School of Cyber Science and Technology, College of Computer Science and Technology, Zhejiang University, Hangzhou, 310027, China, and are also associated with the Key Laboratory of Blockchain and Cyberspace Governance of Zhejiang Province, 330000, China. R. Ma is with Zhejiang Gongshang University. R. Liu is affiliated with Beijing University of Posts and Telecommunications, China

Y. Wang and H. Xu are co-corresponding author.
}
}

\maketitle
\begin{abstract}
In evolving cyber landscapes, the detection of malicious URLs calls for cooperation and knowledge sharing across domains. However, collaboration is often hindered by concerns over privacy and business sensitivities. Federated learning addresses these issues by enabling multi-clients collaboration without direct data exchange. Unfortunately, if highly expressive Transformer models are used, clients may face intolerable computational burdens, and the exchange of weights could quickly deplete network bandwidth. In this paper, we propose Fed-{\scshape urlBERT}, a federated URL pre-trained model designed to address both privacy concerns and the need for cross-domain collaboration in cybersecurity. Fed-{\scshape urlBERT} leverages split learning to divide the pre-training model into client and server part, so that the client part takes up less extensive computation resources and bandwidth. Our appraoch achieves performance comparable to centralized model under both independently and identically distributed (IID) and two non-IID data scenarios. Significantly, our federated model shows about an 7\% decrease in the FPR compared to the centralized model. Additionally, we implement an adaptive local aggregation strategy that mitigates heterogeneity among clients, demonstrating promising performance improvements. Overall, our study validates the applicability of the proposed Transformer federated learning for URL threat analysis, establishing a foundation for real-world collaborative cybersecurity efforts. The source code is accessible at \url{https://github.com/Davidup1/FedURLBERT}.

\end{abstract}


\section{Introduction}

In the constantly evolving landscape of cyber threats, detecting and mitigating malicious URLs has emerged as a pivotal challenge in cybersecurity. Serving frequently as gateways for phishing, malware distribution, and various forms of cyber fraud, malicious URLs pose substantial risks to both individuals and organizations. According to reports from the Anti-Phishing Working Group (APWG), the volume of phishing attacks alone has seen an annual increase of over 150\% since early 2019 \cite{apwgreport}.

Conventional cyber threat countermeasures rely on centralized data acquisition and analysis, typically managed by data centers in a single organization \cite{tajaddodianfar2020texception,li2020improving}. However, this approach may not capture specific threat patterns across various industries or regions and struggles with efficiency and scalability. The need for collaborative efforts in malicious URL detection is evident, but such cooperation is often limited by corporations' reluctance to share data, mainly due to concerns over confidentiality, privacy, and business-sensitive information.

Given these barriers, using a federated learning framework for cross-organizational collaboration is seen as a promising solution. Federated learning \cite{khan2021federated} enables multiple organizations to collaboratively train models for detecting malicious URLs while keeping their data private. By training models locally and sharing only model parameters, not raw data, it addresses concerns about privacy and data security. Moreover, federated learning's adaptability to different technical capabilities and infrastructures among organizations allows for wider cooperation and more effective malicious URL detection.

Pre-trained language models utilizing Transformer architectures \cite{devlin2018bert} have significantly advanced various fields,  including natural language processing\cite{min2023recent}, computer vision\cite{chen2021pre,liu2022video}, and software engineering\cite{feng2020codebert}. Recent studies \cite{haynes2021lightweight,liu2023malicious} have also demonstrated their potential in detecting malicious URLs, indicating the effectiveness of Transformers and pre-training in URL analysis. However, pre-training requires large amounts of data, which can lead to high computational costs and time investment. Acquiring sufficient data and computational resources remains a significant challenge.

Therefore, developing federated pre-trained Transformers presents significant application prospects. This paper introduces a federated pre-training architecture designed for malicious URL detection. Our framework involves retraining a URL-specific pre-trained model in a federated manner and fine-tuning it for the malicious URL detection task. Our approach combines the advanced contextual analysis capabilities of pre-trained models with the efficiency of federated learning in handling distributed data. This method allows multiple participants to collaborate, utilizing their individual data resources to enhance model performance without the need for centralized storage or processing. Such an approach is expected to improve the model's generalizability while accelerating adaptation to newly emerging malicious URL patterns.

The highlights and contributions of our work are as follows:
\begin{itemize}

\item We present a federated, pre-trained Transformer model for malicious URL detection, combining the benefits of federated learning and pre-trained Transformers. It facilitates cross-organizational cooperation and allows worry-free resource contribution. To our knowledge, this is the first federated framework for malicious URL detection, and the first federated pre-trained model for URLs.
\item For practical reasons, we propose a client-side lightweight federated strategy based on split learning and federated learning. This approach avoids full-scale model training on client devices, reducing their computational burden and communication overhead and allowing more participants to join.
\item Our federated pre-trained model achieves performance comparable to centralized models in detecting malicious URLs, even decreasing the FPR by about 7\%. 
\item For the first time, we incorporate the adaptive local aggregation strategy into federated pre-training model. We observe that this strategy shows potential in reducing statistical heterogeneity among federated clients and improving the model's resilience to diverse data distributions.
\end{itemize}

\section{Related Work} 
In this section, we begin with a review of related work in malicious URL detection, followed by an overview of the technical background relevant to our research.

\subsection{Malicious URL Detection}
Malicious URL detection has a long research history, evolving from rule-based approaches to traditional methods based on manual feature engineering, and most recently, to end-to-end deep learning-based methods. While previous work laid the foundation for malicious URL detection research, our focus is primarily on the currently prevalent pre-trained models. Hence, we primarily review studies based on the Transformer architecture. 

Haynes et al.\cite{haynes2021lightweight} apply two state-of-the-art deep transformers, BERT\cite{devlin2018bert} and ELECTRA\cite{clark2020electra}, for phishing detection, utilizing both custom and standard vocabularies. Their research indicates that BERT, pretrained on English text, exhibits strong performance in direct application to URL classification. The literature \cite{10172235} utilizes tiny-Bert, a lightweight version of BERT, to extract URL embeddings for IoT edge malicious URL detection, achieving 99\% accuracy.
\cite{maneriker2021urltran} proposed URLTran, employing transformers, significantly outperforms other deep learning methods in phishing URL detection with low FPRs, achieving a TPR of 86.80\% at an FPR of 0.01\%, and maintaining robustness against classical adversarial phishing attacks. The ref.\cite{10095719} trained a specialized tokenizer for URL data, adjusted the pre-training task of the BERT model and proposed PhishBERT, achieving a TPR increase of 7\% and 25\% respectively over URLTran on different datasets, while maintaining an extremely low FPR. The ref.\cite{c} employs BERT for feature extraction from URL data and utilized a deep learning network in downstream tasks related to phishing URL detection, achieving an accuracy of 96.66\% and significantly outperforming the baseline established by traditional machine learning methods across various metrics. Past studies have substantially confirmed the validity of pre-trained models for malicious URL detection. Our exploration of federated URL pre-trained models combines the advantages of federated learning and pre-training, expanding data access and allowing users with less computational power to benefit from sophisticated models.

\subsection{Technical Background}
Our study is founded on three critical technological pillars: the classic pre-trained model BERT\cite{devlin2018bert}, federated learning, and split learning. Here, we introduce the fundamentals of these three technologies.
\subsubsection{BERT}
In our research, we utilize the BERT architecture for pre-training a URL-specific model. Therefore, we now review the background of BERT.

BERT is a groundbreaking language model pre-trained on a stack of Transformer encoders \cite{DBLP:journals/corr/abs-1810-04805}. The Transformer block consists of two sublayers: a multi-head attention layer and a fully connected feed-forward neural network layer. The encoder applies residual connections around each sublayer. BERT introduces an optimized approach by defining two training objectives -- the Masked Language Model (MLM) and Next Sentence Prediction (NSP). These objectives empower the pre-training model to learn in a self-supervised manner from a large scale dataset, fostering a profound comprehension of the intricate data structure and inherent linguistic phenomena. We omit detailed elaboration on the NSP task due to its minimal impact on performance as demonstrated in previous studies \cite{10.1109/MILCOM52596.2021.9653028} and its irrelevance in our URL pre-training process.

\textbf{Masked Language Model:} The MLM is a task that involves predicting the original tokens from a modified input where some of the tokens have been replaced. Specifically, a subset of tokens $Y \subseteq X$ is selected from a token sequence $X$ and replaced with different tokens. In the implementation of BERT, $Y$ accounts for 15\% of the tokens in $X$. Among these tokens, 80\% are replaced with the [MASK] token, 10\% are replaced with randomly selected tokens based on the unigram distribution, and the remaining 10\% are left unchanged. BERT independently selects each token in $Y$ by randomly choosing a subset. In our research, we also employ MLM as the pre-training objective for URLs.During the fine-tuning phase, typically incorporating a fully connected layer into the BERT model expedites its adaptation to downstream tasks, leading to peak performance within a limited number of training epochs.

\subsubsection{Federated Learning}
Federated learning, a concept pioneered by Google in 2017 \cite{pmlr-v54-mcmahan17a}, marks a significant shift in the machine learning landscape. It facilitates the distributed training of AI models across multiple remote computing nodes, eliminating the necessity for direct data transfer. This methodology proves particularly vital in contexts where stringent privacy and data security are paramount, such as in healthcare\cite{rieke2020future}, finance\cite{li2020federated}, or government sectors\cite{li2020review,}. Here, data sovereignty is preserved, with the data remaining securely on its native servers. Under this paradigm, the focus is on exchanging model parameters or training insights rather than the actual data, thereby upholding data privacy. Federated learning's introduction has been a cornerstone in addressing data privacy challenges, enhancing collaborative efforts across organizations, and promoting the sharing of insights and knowledge.
\begin{figure}
    \centering
    \includegraphics[width=2.5in,height=2.3in]{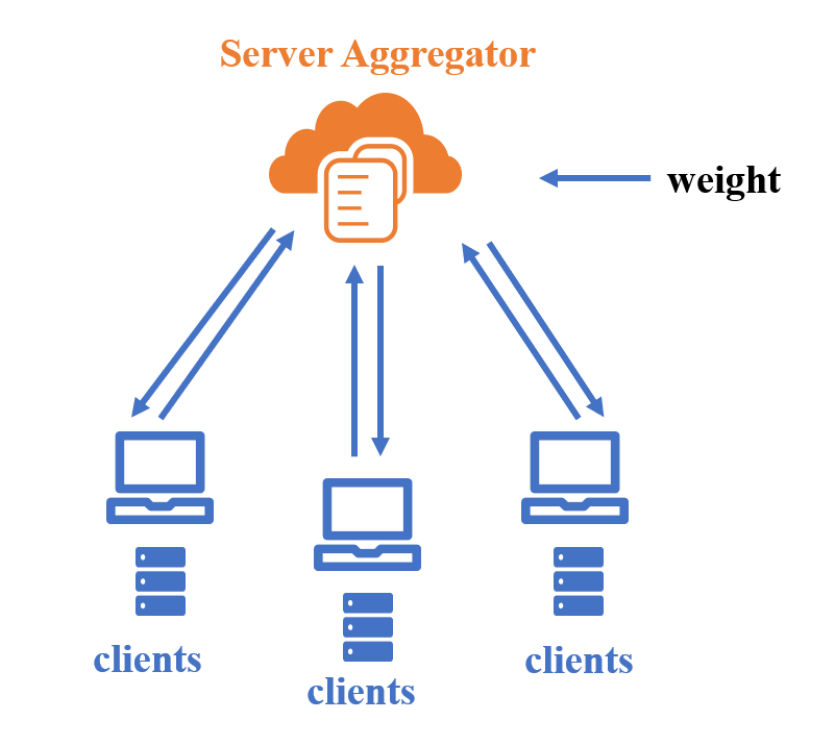}
    \caption{The classical architecture of the FedAvg algorithm}
    \label{fig:architecture}
\end{figure}

The classic federated architecture is illustrated in Fig. \ref{fig:architecture}. In federated learning under the \textit{client-server} architecture, the model
training process is performed locally on the client side while the
generated gradient values or model weights are sent to the server. The
server is responsible for aggregating them to form a global model, which
is then distributed to each client for the next epoch of training. The
fundamental characteristic of federated learning lies in replacing data
exchange with the exchange of model weights or gradient values.
Currently, one of the most effective strategies for model aggregation in
federated learning is FedAvg \cite{pmlr-v54-mcmahan17a} and one of the most prominent areas
of research in the academic community focuses on the performance and
improvement of federated learning methods under Non-iid data
distributions, such as FedALA \cite{Zhang_Hua_Wang_Song_Xue_Ma_Guan_2023} and FedProx \cite{li2020federated}.
Traditional federated learning deploys model training on the local clients. As the parameter size of the trained model increases, an inevitable consequence arises wherein an increasing number of clients become unable to bear the computational and storage costs associated with model training. This phenomenon undermines client participation in federated training, consequently leading to a partial loss in data richness, as has been demonstrated in previous studies \cite{alam2022fedrolex,thapa2022splitfed}. Our study highlights the unsuitability of traditional client-based federated learning methods for large-scale, pre-trained machine learning models. In response, we propose an alternative, more efficient client learning approach.

\subsubsection{Split Learning}

\begin{figure}
    \centering
    \includegraphics[width=3.7in,height=2.59in]{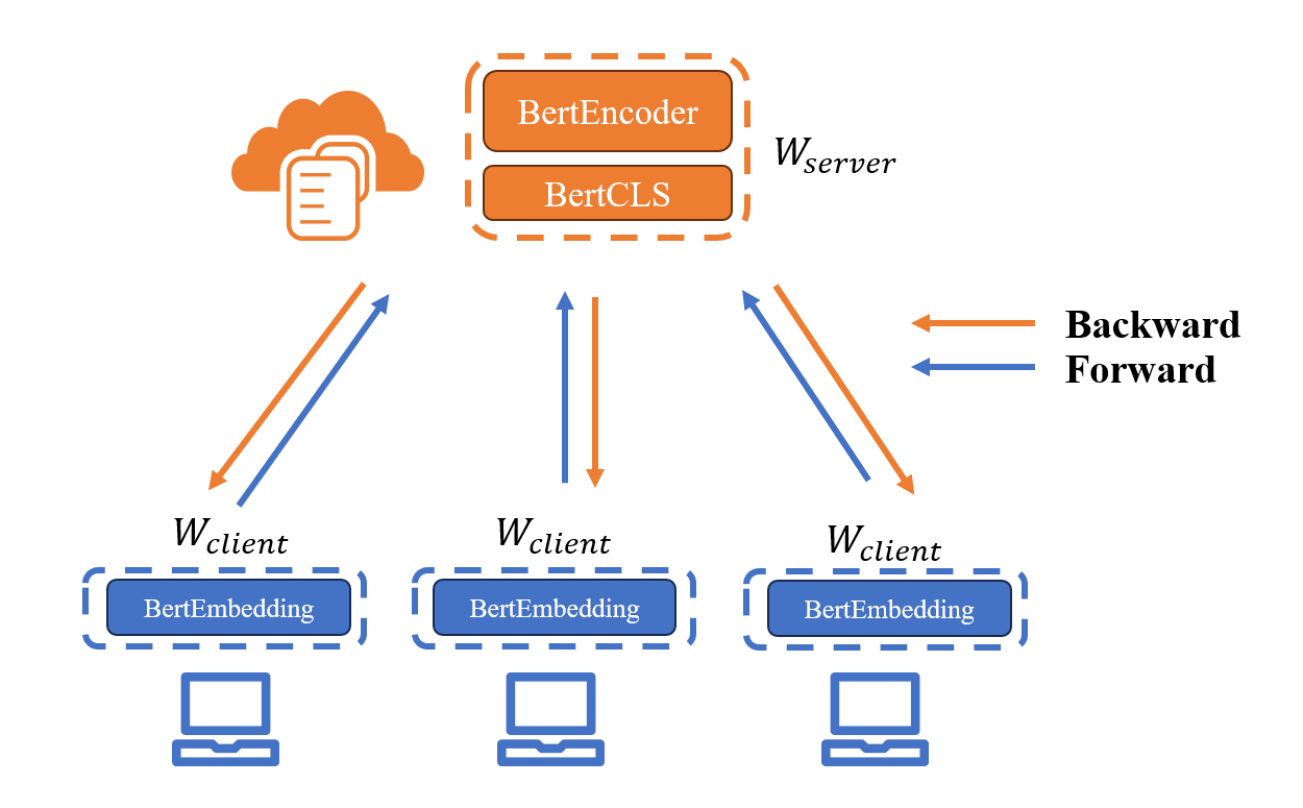}
    \caption{The decomposition approach of Split Learning for pre-trained
models}
    \label{fig:split}
\end{figure}

Split learning \cite{gupta2018distributed} is a machine learning strategy which dissects a complex neural network into multiple segments, handled and computed across various devices. The scenarios for multiple clients can be based on multiple regular computing nodes (Alices) + one high
computing node (Bob) \cite{thapa2022splitfed,tian2022fedbert}. In this strategy, each client
trains a complete model together without disclosing the original data,
while allocating the computation-intensive part of the model to the high
computing node. In split learning, the model is usually divided into two
parts. Alices perform forward propagation on their local data and
transmit the intermediate results to Bob. Bob then continues with
forward and backward propagation, and sends back the intermediate data
of backward propagation to Alices, as shown in Fig. \ref{fig:split}. However, the coordination of the learning process among multiple clients occurs through either a centralized mode or a peer-to-peer mode within Split Learning \cite{gupta2018distributed}, leading to high training time cost.

Previous studies \cite{tian2022fedbert} have established the effectiveness of combining split learning with federated learning. In this work, we adopt a similar concept but utilize a different split strategy to build our Fed-{\scshape urlBERT}. In our approach, the BertEmbedding layer is located on the client side, while the Transformer layers are deployed on high-performance servers.

\section{Fed-{\scshape urlBERT}} 

In this section, we describe the overall framework of Fed-{\scshape urlBERT}, followed by a detailed discussion on the implementation of federated learning in both the pre-training and fine-tuning stages, utilizing decentralized data.

\subsection{Framework}

\begin{figure*}
    \centering
    \includegraphics[scale=0.35]{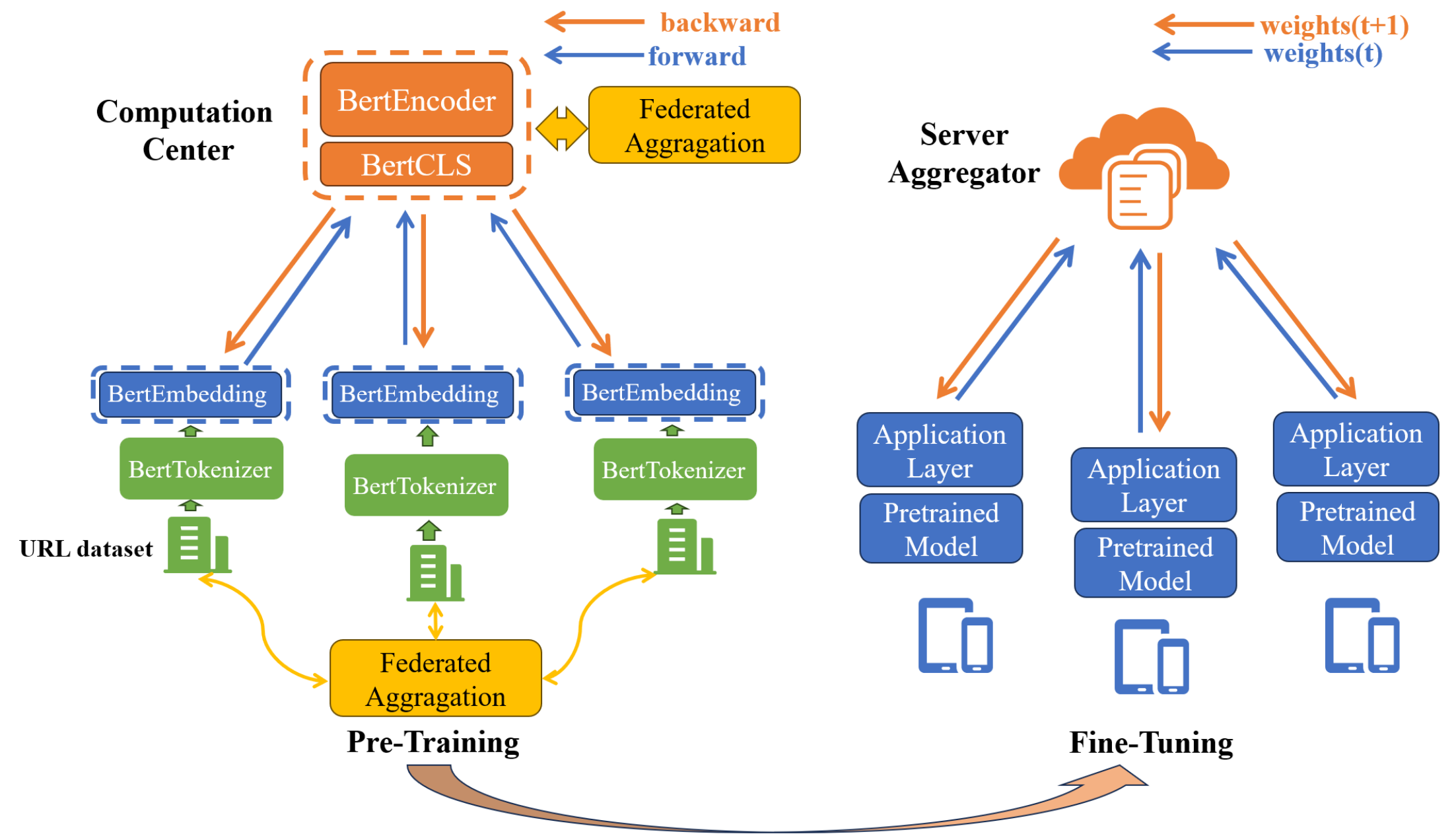}
    \caption{Architecture diagram of Fed-{\scshape urlBERT}}
    \label{fig:Framework}
\end{figure*}

The architecture of Fed-\textsc{urlBERT}, as depicted in Fig. \ref{fig:Framework}, involves pre-training and fine-tuning a BERT model on URL data using federated learning and split learning principles. Our architecture adopts a client-light approach, where a computational center handles the intensive learning tasks of the BERT encoder, while federated participants focus on locally training the Tokenizer and encoding their own data. During the pre-training stage, the Bert model \(W\) is partitioned into \(W_{client}\) and \(W_{server}\). Both the computational center server and the federated participants engage in forward and backward parameter updates through federated communication. Specific details of the training are provided in Section \ref{section2}. In the fine-tuning phase, the application layer of BERT is distributed among all federated participants, used for learning the malicious URL detection task on the federated client side. Parameter aggregation is conducted using the Federated Averaging Algorithm, FedAvg, as detailed in Section \ref{section3}.

\subsection{Pre-Training}
\label{section2}
\begin{figure}[t]
    \centering    \includegraphics[width=3.6in,height=2.1in]{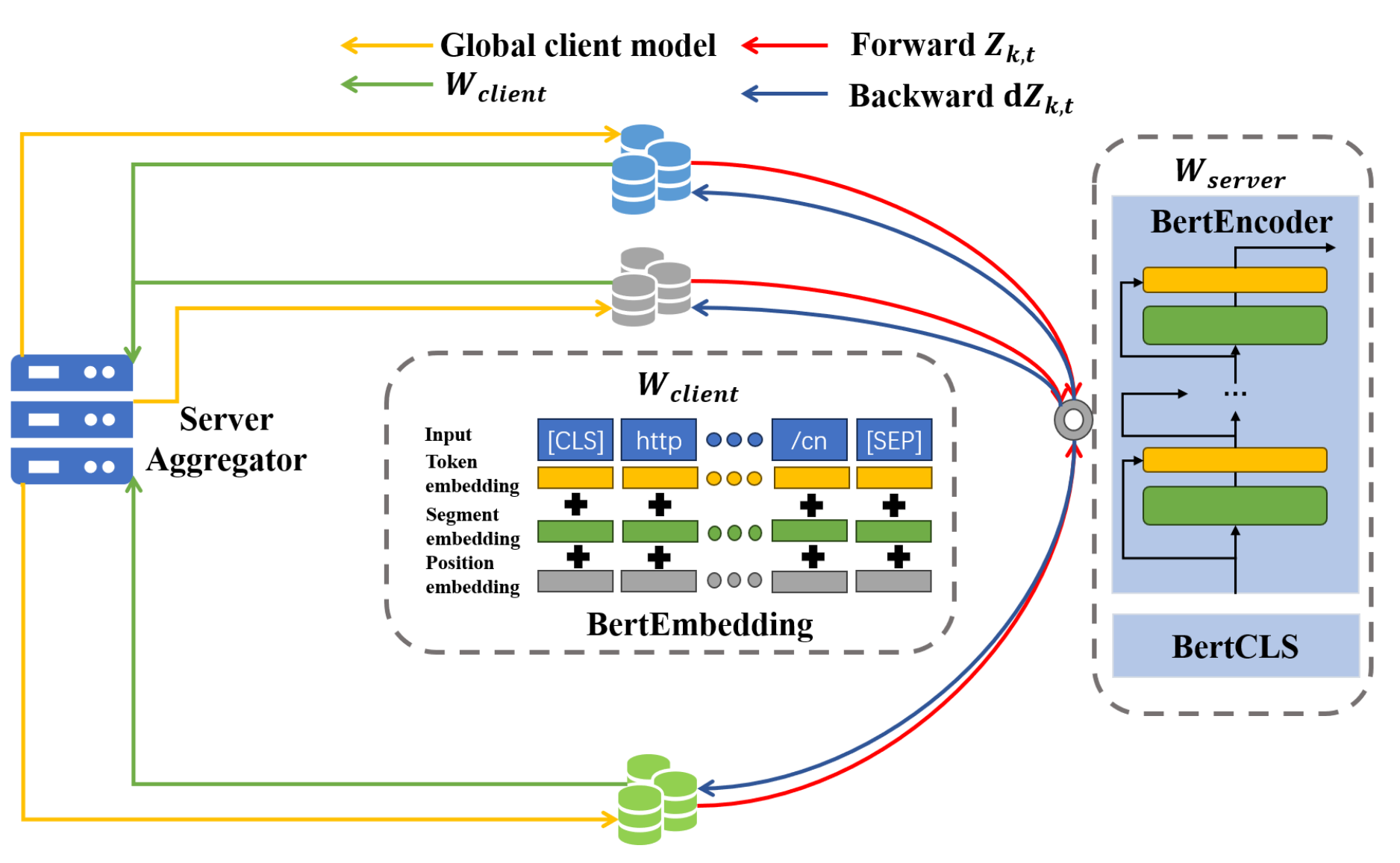}
    \caption{A schematic illustration of the integration of Federated Learning
and Split Learning}
    \label{fig:SFL}
\end{figure}

We pre-train URLs using the Masked Language Model (MLM) as our training objective. In the MLM task, \(15\%\) of the tokens are replaced with [MASK], of which \(10\%\) remain unchanged, and another \(10\%\) are randomly substituted with a token from the vocabulary. We split the BERT model into \(W_{\text{client}}\) and \(W_{\text{server}}\) based on split learning principles, and aggregate federated parameters using the FedAvg algorithm, as illustrated in Fig. \ref{fig:SFL}.

The BertEmbedding network of the BERT model is allocated to \(W_{\text{client}}\), while the BertEncoder (the Transformer's encoder component), which houses a larger parameter set and incurs substantial computational costs during training, along with the BertOnlyMLMHead network layer, is assigned to the computation center’s \(W_{\text{server}}\). Selected clients feed their local data into their respective local models concurrently. The intermediate outputs from the local model’s forward propagation, denoted as \(Z_{k,t}\), are transmitted from the clients to the server. The forward propagation then continues on the server's network, \(W_{\text{server},k,t}\), yielding an output. This is followed by backward propagation, where \(dZ_{\text{server},k,t}\) is sent back to \(W_{\text{client}}\) for additional backward propagation steps. This cycle completes one training iteration, as depicted in Algorithm \ref{alg:pre-training}.

\begin{algorithm}[h]
    \caption{Pre-Training of FedPhishBert}
    \textbf{Notations:} (1)\(S_{t}\) is a set of \(K\) randomly selected
clients at t-th round, (2) \(Z_{k,t}\ \)is the intermediate value of the
forward propagation, (3) \({Y_{k}}\) and \(\widehat{Y_{k}}\) respectively represent tensors of the original words marked with [MASK] obtained through tokenization based on our custom vocab, and the model's inferred outputs on client k, (4) \(loss_{k}\) is the loss of the client k, (5) \(n\) and \(n_{k}\) are
the total sample size and sample size at client k, respectively, (6)
\(\mathbf{\eta}\) is the learning rate.

    \label{alg:pre-training}
    \begin{algorithmic}[1]
        \small
        \Function{ComputationServerExecutes}{$Y_{k},Z_{k,t}$}
            \If{t=0}
                \State Initialize \(W_{server,t}\)(global model at server)
            \Else
                \For{each Client k  \(\in S_{t}\) in parallel}
                    \While{local epoch e \(\neq E\)}
                        \State $Z_{k,t},Y_{k} \mathbf{\leftarrow} ClientUpate(W_{client,k,t})$
                        \State Forward propagation, and compute  $\widehat{Y_{k}}$
                        \State Calculate $loss_{k}$ with $\widehat{Y_{k}}$ and $Y_{k}$
                        \State Backward propagation and calculate $dZ_{k,t}$
                        \State Send $dZ_{k,t}$ to Client k for ClientBackProp
                    \EndWhile
                \EndFor
                \State Server model update:
                \State $W_{s,t+1}\leftarrow W_{s, t}-\eta\frac{n_{k}}{n}\sum_{i=1}^{K} \nabla \operatorname{loss}_{i}\left(\boldsymbol{W}_{s, i, t} ; Z_{i, t}\right)$
                 
                 ($W_{s}$ is the abbreviation of $W_{server}$)
            \EndIf
        \EndFunction

        \Function{FedServerExecutes}{$W_{server,k,t}$}
            \If{t=0}
                \State Initialize $W_{client,t}$ (global client model)
                \State Send $W_{client,t}$ to all clients for ClientUpdate$(W_{client,k,t})$
            \Else
                \For{each client k $\in S_{t}$ in parallel}
                    \State $W_{client,k,t}\leftarrow ClientBackProp(dZ_{k,t})$
                \EndFor
                \State global client model updates:
                \State $W_{client,t + 1} \leftarrow \sum_{k = 1}^{K}\frac{n_{k}}{n}W_{client,k,t}$
                \State Send $W_{client,t + 1}$ to all clients
            \EndIf
        \EndFunction
    \end{algorithmic}
\end{algorithm}

After reaching the specified number of training rounds, the model
aggregation server applies the FedAvg scheme to aggregate \(W_{client}\)
and redistributes it to each client. The computation center will also
aggregate \(W_{server}\).

The \(W_{server,k,t}\) here represents the server-side network in the computation center for client k in the t-th round of training. To enable
parallel execution of client training and communication, a copy of
\(W_{client}\) is set for each client on the server side. Once the
specified number of training epochs is completed, the global model is
aggregated and deployed to each copy. \(Z_{k,t}\) represents the
intermediate data propagated by client k in the t-th round of training.
This training approach not only ensures the privacy and security of
client data but also reduces computational costs for clients with
limited resources. Furthermore, it enhances communication efficiency
during the model training process and improves the robustness of the
trained model.

\subsection{Fine-Tuning}
\label{section3}
In the fine-tuning stage, a fully connected layer is added to the top of the model on each client enable binary classification of URLs. The model undergoes local training using fine-tuning data on individual clients, a phase that demands relatively lower computational resources and is feasibly managed by each participant. This efficiency validates the use of the classical federated learning approach \cite{pmlr-v54-mcmahan17a}. Fine-tuning of the complete model occurs locally at each client, involving the exchange of model parameters between clients and the server. Parameter aggregation is executed at the server, as delineated in Algorithm \ref{alg:FedAvg}.

\begin{algorithm}[h]
    \caption{Implementation of Fine-Tuning}
    \textbf{Notations:} (1)\(S_{t}\) is a set of \(K\) randomly selected
clients at t-th round,(2)bz is the local batch size,(3) \(\mathbf{\eta}\) is the learning rate,(4) \(Data_{k}\) is the local data at \(client_{k}\)

    \label{alg:FedAvg}
    \begin{algorithmic}[1]
        \small
        \Function{ServerExecutes}{$W_{k,t}$}
            \If{t=0}
                \State Initialize \(W_{0}\)(global model)
                \State Send $W_{0}$ to all clients
            \Else
                \For{each epoch}
                    \For{each client $k\in S_{t} $in parallel}
                        \State $W_{k,t+1} \mathbf{\leftarrow} ClientUpdate(k,W_{k,t})$
                    \EndFor
                    \State $\boldsymbol{W}_{t+1} \leftarrow \sum_{k=1}^{K} \frac{n_{k}}{n} \boldsymbol{W}_{k, t+1}$
                    \State Send $W_{t+1}$ to all Clients
                \EndFor
            \EndIf
        \EndFunction

        \Function{ClientUpdate}{$k, W_{k,t}$}
            \State $B \leftarrow\left(\text { split } \text { Data }_{k} \text { into batches of size bz }\right)$

            (In the subsequent improvements, local initialization will also be conducted)
            \For{each local epoch}
                \For{batch $b \in B$}
                    \State $W \leftarrow W-\eta \nabla \operatorname{loss}(W ; b)$
                \EndFor
            \EndFor
            \State return $W$ to server
        \EndFunction
    \end{algorithmic}
\end{algorithm}

In each training round, a random subset of \(n\) clients is chosen for training. After these clients complete their local training, their local models, denoted as \(\mathbf{W}_{k,t}\), are sent to the server to update the global model. Once all participating clients finish their training, the server aggregates the model parameters using a weighted averaging method, then redistributes the updated model to all clients for the next training epoch. \[w_{t + 1} \leftarrow \sum_{k = 1}^{K}\frac{n_{k}}{n}w_{k,t}\] Here, \(\mathbf{W}_{k,t}\) represents the model of client \(k\) in the \(t\)-th training round, while \(w_{t + 1}\) signifies the aggregated model from the \(t\)-th round, serving as the global model for the ensuing round.

Additionally, considering that federated learning faces the challenge of non-independent and non-identically distributed (non-i.i.d) data \cite{arnold2021demystifying,zhu2021federated}, characterized by dissimilar features and distributions across different datasets, where feature interdependencies are not mutually independent. These distributional disparities among datasets can lead to a reduction in the predictive performance of the globally aggregated model on individual client data and have adverse effects on model convergence \cite{sattler2019robust, lifederated}. Previous efforts have been dedicated to enhancing federated learning models under non-IID data settings. Taking inspiration from FedALA \cite{Zhang_Hua_Wang_Song_Xue_Ma_Guan_2023}, we introduce an initialization step before each client's training session. Unlike conventional federated learning, which starts each training round by overwriting the client model with the global model, we merge the global model with the client model using a specific weight \(K_{i}^{t}\), where these weights \(K_{i}^{t}\) are determined through a learning process.

It is important to note that, unlike FedALA \cite{Zhang_Hua_Wang_Song_Xue_Ma_Guan_2023}, considering the possibility of further parameter freezing on the BERT model in subsequent work and the differences between the targeted model network structures, we have included all model parameters involved in the training process in the local initialization.

\[\widehat{W_{i}^{t + 1}} = W_{i}^{t} + \left( W^{t} - W_{i}^{t} \right) \odot K_{i}^{t}\]

During the initialization phase, in consideration of the computational cost at the client-side, a portion of the client's training data is strategically allocated for model training. The model undergoes updates by iteratively adjusting the parameter \(K\) until the variance of the loss predicted by the model falls below a predefined threshold, leading to the termination of the initialization process. This approach aims to tailor the client's training model to better match the local data distribution of the client, thereby enhancing the generalization and robustness of the globally aggregated model. We investigate the impact of this approach on model training under three different data distribution scenarios:

\begin{enumerate}
  \item Training and testing data are both independent and identically distributed (i.i.d.), referred to as IID.
  \item Training data is non-i.i.d., while testing data is i.i.d., denoted as Non-IID(2).
  \item Both training and testing data follow the same distribution but are not independently and identically distributed, labeled as Non-IID(3).

\end{enumerate}

\section{Experiments} 

In this section, we evaluate the performance of models within a federated architecture. Our experimental objectives are as follows:
\begin{enumerate}
    \item Comparing between federated and central training models in phishing URL detection.
    \item Assessing the effect of data distribution in fine-tuning on federated models.
    \item Evaluating FedALA-based fine-tuning across different data scenarios.
    \item Validating client proportion selection per training round in fine-tuning.
    \item Analyzing the impact of parameter freezing in fine-tuning on federated outcomes.
\end{enumerate}

\begin{figure}
    \centering
    \includegraphics[scale=0.3]{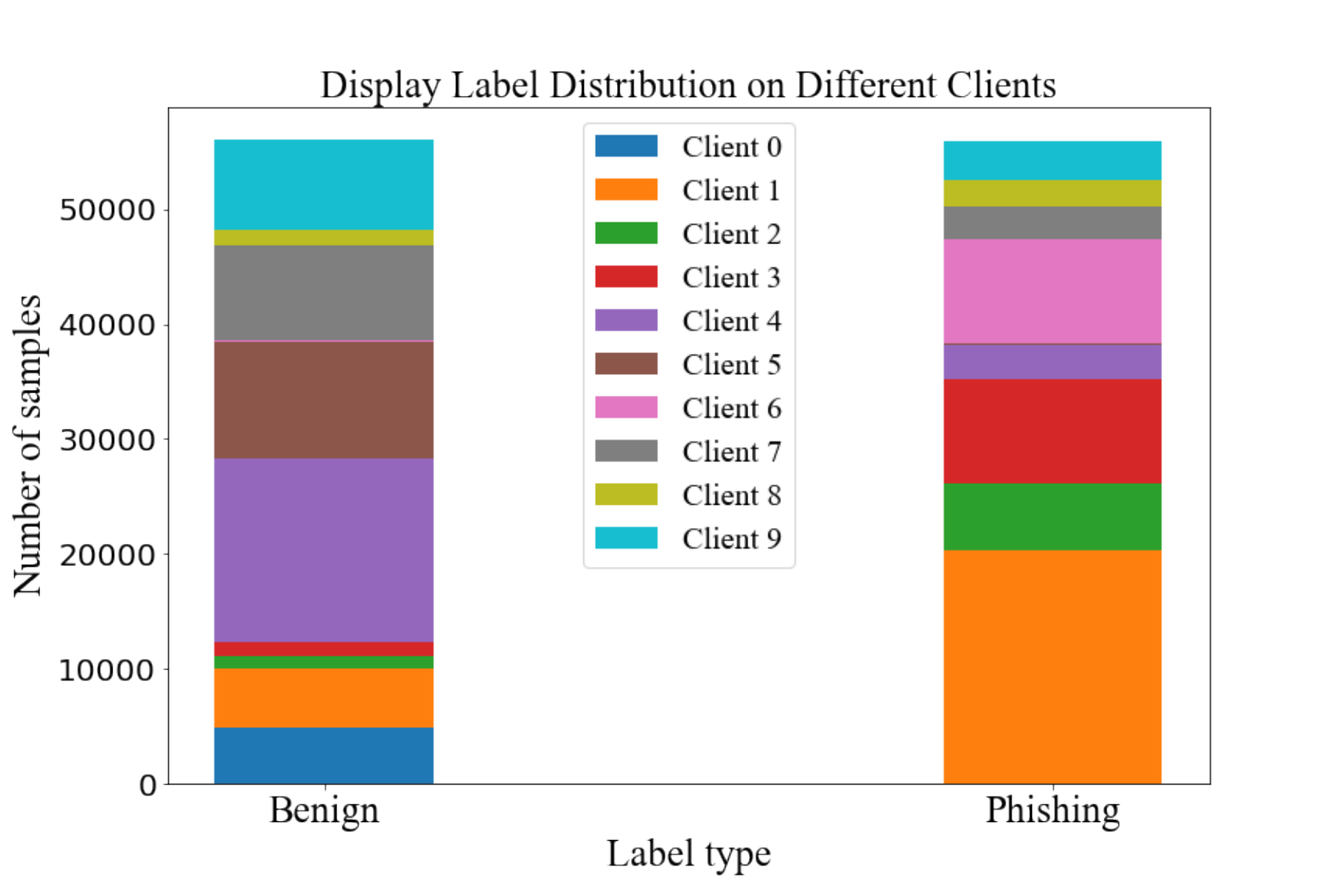}
    \caption{\textbf{Lable distribution with} \(\mathbf{\alpha}\) \textbf{value
of 0.7}}
    \label{fig:distribution}
\end{figure}

Regarding data distribution, we set up three scenarios: both training and testing data following an i.i.d. pattern; only testing data is i.i.d., with training data following a Dirichlet distribution (\(\alpha\)=0.7); and both training and testing data derived from Dirichlet distributions with an \(\alpha\) value of 0.7. Fig. \ref{fig:distribution} illustrates the data distribution at this stage.

\subsection{Dataset}
Previous studies \cite{liu2023malicious,10095719} have already demonstrated the significant effectiveness of pre-trained URL language models. In our research, we do not seek to reconfirm this but instead focus on and aim to develop a federated pre-training model scheme specifically for the URL data domain. Hence, our experiments mainly evaluate the performance differences between federated and central pre-training. Under this premise, we use a relatively small dataset, the GramBeddings Dataset, consisting of 799,992 URLs provided by Grambeddings \cite{10.1016/j.cose.2022.102964}, for training and fine-tuning our URL language model. This dataset is evenly split with 399,992 malicious and 400,000 benign URLs. Malicious URLs were sourced from PhishTank and OpenPhish, collected between May 2019 and June 2021. Benign URLs were gathered through web crawling of popular sites listed in Alexa, using down-sampling to maintain balance. Among these, 52.17\% of benign URLs used the \textit{.com} TLD, 12.04\% were ccTLDs, and 35.79\% other gTLDs. The malicious URLs consisted of 60.10\% \textit{.com}, 11.82\% ccTLDs, and 28.08\% other gTLDs. For our study, 639,994 data points are utilized for pre-training the model, while the remaining 159,998 are used for fine-tuning.

\subsection{Infrastructure}
\textbf{Model Construction.} We configure the BERT model for pre-training using HuggingFace's Transformers interface, tailoring it to our requirements. Initially, we set the \texttt{vocab\_size} to 1000, crafting a specialized vocabulary for our URL dataset. We then reshape the model's embedding matrix according to the \texttt{vocab\_size}, configuring it to (1000, 768). We utilize the GELU activation function\cite{hendrycks2016gaussian} and set dropout rates for the attention and hidden layers at 0.1 and 0.2, respectively.

\noindent\textbf{Pre-Training.} The pre-training of our model is conducted on a single NVIDIA A100 (80GB) GPU across 10 training epochs. We use the Adam optimizer with a learning rate of 5e-5 and a batch size of 64. During each epoch, all clients are engaged in the training, with both client and server-side training iterations set consistently to 1.

\noindent\textbf{Fine-Tuning.} We employ a single NVIDIA GeForce RTX 3090 (24GB) GPU for fine-tuning, using the Adam optimizer with a learning rate of 2e-6 and a batch size of 32. We configure 10 clients in total, randomly selecting 5 for training in each epoch. Each client conducts 5 local training iterations per epoch, over a total of 30 epochs.

\subsection{Result}
\subsubsection{Comparasion with Centralized Training}
Our study develops a federated learning-based training scheme for employing Transformer in phishing link detection. We compare centralized and federated training models across key metrics, with FedAvg(*) denoting our method of local initialization during fine-tuning to enhance training efficiency. This experiment tests federated learning's effectiveness for phishing detection and contrasts it with traditional centralized training. Our datasets consisted of 639,994 records for pre-training and 159,998 for fine-tuning, uniformly used across all experiments. In the federated learning group, the dataset distribution to clients followed an independent and identically distributed (i.i.d.) manner\cite{zhao2018federated}.
\begin{table}[t]
\caption{\textbf{Comparison of various metrics of models obtained after 30 training epochs under independently and identically distributed (i.i.d.) data, where FedAvg(*) denotes the fine-tuning approach improved by local initialization.}}
    \centering
    \renewcommand{\arraystretch}{1.5} \resizebox{0.46\textwidth}{!}{
    \begin{tabular}{lccccc}
    \hline
    & ACC & TPR & FPR & F1-SCORE & AUC \\
    \hline
    Centralized & \textbf{95.31\%} & \textbf{0.9571} & 0.0510 & \textbf{0.9516} & \textbf{0.9530} \\
    FedAvg & 94.73\% & 0.9386 & \textbf{0.0438} & 0.9450 & 0.9474 \\
    FedAvg(*) & 94.82\% & 0.9416 & 0.0449 & 0.9461 & 0.9483 \\
    \hline
    \end{tabular}}
\label{tab:tab1}
\end{table}

\begin{table}[t]
\caption{\textbf{Accuracy comparison of the model under three experimental scenarios with the centrally trained model. (2) represents the second data distribution setting, and (3) represents the third. The accuracy value is obtained by computing the arithmetic mean of the test accuracies across all client clients.}}
    \centering
    \renewcommand{\arraystretch}{1.5} 
    \resizebox{0.46\textwidth}{!}{
    \begin{tabular}{lcccc}
    \hline
    & Centralized & IID & Non-IID(2) & Non-IID(3) \\
    \hline
    ACC & \textbf{95.31\%} & 94.73\% & 94.64\% & 94.18\% \\
    \hline
    \end{tabular}}
\label{tab:tab2}
\end{table}

\begin{table}[t]
\caption{\textbf{Comparison of various metrics for models fine-tuned under FedAvg in three distinct experimental scenarios.}}
\centering
\renewcommand{\arraystretch}{1.5}
\resizebox{0.46\textwidth}{!}{
\begin{tabular}{lccccc}
\hline
& ACC & TPR & FPR & F1-SCORE & AUC \\
\hline
IID & \textbf{94.73\%} & 0.9386 & \textbf{0.0438} & \textbf{0.9450} & \textbf{0.9474} \\
Non-IID(2) & 94.64\% & \textbf{0.9455} & 0.0532 & 0.9446 & 0.9462 \\
Non-IID(3) & 94.18\% & 0.9389 & 0.0588 & 0.9178 & 0.9398 \\
\hline
\end{tabular}}
\label{tab:tab3}
\end{table}

Table \ref{tab:tab1} presents a comparative analysis of experimental results between models obtained through centralized training and federated learning. It is observed that, in contrast to centralized training, the models trained under the proposed methodology demonstrate comparable performance across metrics such as Accuracy, TPR, F1-Score, and AUC. Notably, in FPR, a pivotal metric in the task of URL detection, our proposed training method showcases a noteworthy reduction of about 7\% compared to centralized training, declining from 0.0510 in centralized training to 0.0438 in our proposed approach. Moreover, among these metrics, the most significant disparity observed when compared to the centralized training models lies in the TPR metric, which exhibits a performance decrease of 1.9\% at 0.9386 in contrast to 0.9571 in centralized training. Additionally, we observed that besides the FPR, the improved fine-tuning method led to performance enhancements in the other four evaluation metrics.

This result indicates that the federated training approach adopted in our study notably reduces the false positive rate of the model on benign samples, while ensuring that the performance deviation stays within a maximum of 1.9\% in specified metrics. Such a reduction is vital for enhancing the browsing experience of users in Internet applications.

\subsubsection{Impact of Diverse Data Distribution Scenarios}
In the field of federated learning, a key area of research is exploring how shifts in data distribution affect model training. In practical settings, the training data for models often show characteristics that are not independent and identically distributed (Non-IID). Therefore, it becomes crucial to examine how non-IID data influences the training process. In our experiments, we established three scenarios of data distribution:

\begin{enumerate}
    \item Training and test data on each client adhere to an independently and identically distributed pattern.
    \item Training data on each client follows a Dirichlet distribution, while test data maintain an independent and identically distributed pattern.  (Experimental Group Non-IID(2))
    \item Both training and test data on each client conform to the same Dirichlet distribution. (Experimental Group Non-IID(3))
\end{enumerate}

The second distribution investigates the generalization of models trained under personalized training data on each client when applied to diverse testing data. The third distribution studies the performance of each client when encountering personalized testing data post personalized training. We compare the extent of performance degradation of federated training models compared to centralized training models in terms of the Accuracy. Additionally, we contrast the model performance across three different data distributions in the experimental setup, considering various metrics.

Here, the emphasis lies in delineating the background of our experiments: multiple security agencies are collaboratively engaged in the pre-training of models within federated learning framework and these pre-trained models are deployed across a multitude of user devices to undertake federated fine-tuning under intricate data distributions. Here, we posit that during the pre-training, security agencies, specialized entities in information security, undertake the data processing task to mold the pre-training datasets into an ideal distribution. Thus, our experimental manipulation of diverse data distributions focuses on the model's performance during fine-tuning. The pre-trained BERT models we employ are derived from federated pre-training scenarios where data is independently and identically distributed.

The experimental results in Table \ref{tab:tab2},\ref{tab:tab3} demonstrate the adverse impact of real-world data distributions on models trained through federated learning. Judging by the Accuracy metric, the performance of models trained under non-independent and identically distributed (Non-IID) data exhibits a decline compared to centralized training models, dropping to 94.64\% and 94.18\% in the Non-IID(2) and Non-IID(3) experimental groups respectively. When comparing with the scenario where data follows an independent and identically distributed pattern, models trained on non-IID data show a decrease in performance across Accuracy, FPR, F1-Score, and AUC. However, surprisingly, the model showcases superior performance in terms of TPR in both non-IID scenarios, respectively rising to 0.9455 and 0.9389. Nevertheless, in the worst case, the model's performance decreases by 0.58\% and 0.80\% in Accuracy and AUC metrics, when compared to the IID scenario. This underscores the robustness of FedAvg's scheme in the face of non-IID data distributions. To enhance model performance, drawing inspiration from work of FedALA\cite{Zhang_Hua_Wang_Song_Xue_Ma_Guan_2023}, we refine the model's training methodology.

\begin{figure*}[t]
    \centering
    \includegraphics[scale=0.45]{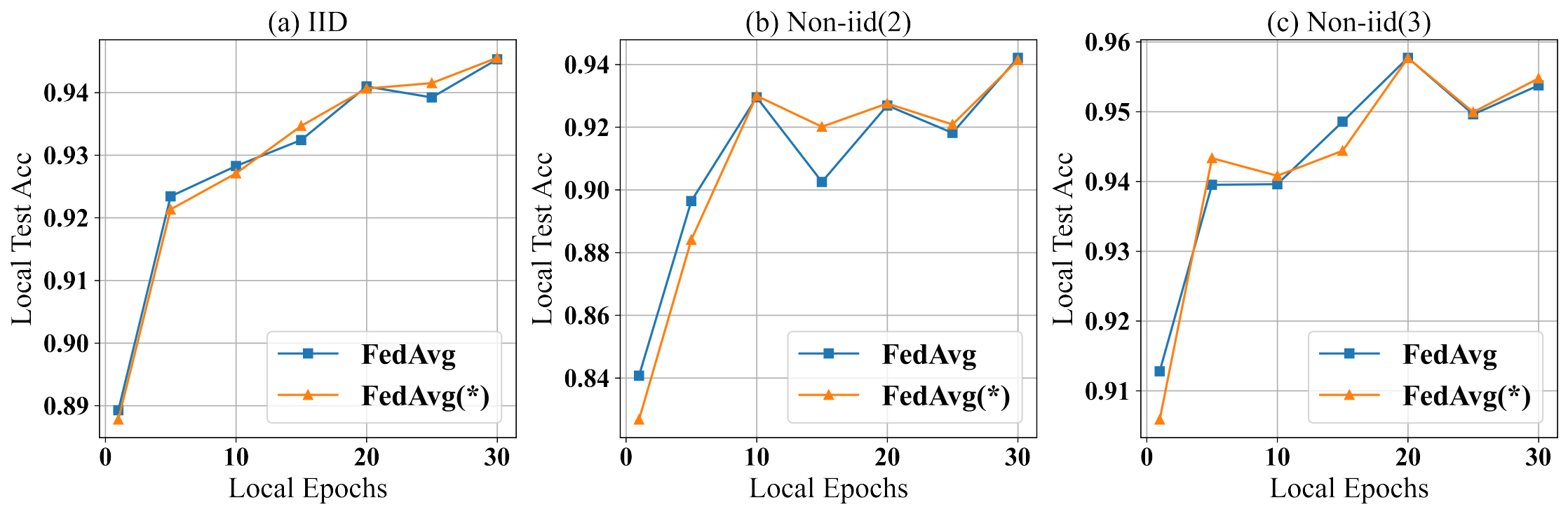}
    \caption{\textbf{Comparison of the accuracy of the models before and after
improvement during the training phase under different data
distributions.}}
    \label{fig:accuracy_comparasion}
\end{figure*}

\begin{table*}[t]
\caption{\textbf{Comparison of the performance of models before and after improvements under three data distributions.}}
\centering
\renewcommand{\arraystretch}{1.5}
\resizebox{0.76\textwidth}{!}{\scriptsize
\begin{tabular}{lcccccc}
\hline
& \multicolumn{2}{c}{IID} & \multicolumn{2}{c}{Non-IID(2)} & \multicolumn{2}{c}{Non-IID(3)} \\ 
\cline{2-7}
& FedAvg & FedAvg(*) & FedAvg & FedAvg(*) & FedAvg & FedAvg(*) \\       
\hline
ACC & 94.73\% & \textbf{94.82\%} & 94.64\% & \textbf{94.74\%} & 94.18\% & \textbf{94.57\%} \\
TPR & 0.9386 & \textbf{0.9416} & 0.9455 & \textbf{0.9520} & 0.9389 & \textbf{0.9488} \\
FPR & \textbf{0.0438} & 0.0499 & \textbf{0.0532} & 0.0579 & \textbf{0.0588} & 0.0618 \\
F1-SCORE & 0.9450 & \textbf{0.9461} & 0.9446 & \textbf{0.9454} & 0.9178 & \textbf{0.9189} \\
AUC & 0.9474 & \textbf{0.9483} & 0.9462 & \textbf{0.9471} & 0.9398 & \textbf{0.9431} \\
\hline
\end{tabular}}
\label{tab:tab4}
\end{table*}

\subsubsection{Evaluation of FedALA}
We investigate the impact of our modification to the FedAvg algorithm on model training with these data distributions, utilizing the module ALA \cite{fedala}. Thanks to the remarkable portability and modular characteristics inherent in the collaborative work of FedALA \cite{Zhang_Hua_Wang_Song_Xue_Ma_Guan_2023}, despite differences in specific training methodologies, we were able to seamlessly integrate the ALA module \cite{fedala} into our training scheme. Through minor adjustments, we tailored the ALA module to suit our experimental setup. In this experiment, we applied various methodologies to preprocess the same dataset and allocated it to multiple clients, conducted training separately under the FedAvg framework and the enhanced approach, with comparative analysis performed to observe the model's performance at different temporal checkpoints during the training process. This investigation aimed to delve into the effectiveness of the enhanced approach concerning improvements in model performance and optimization of the training procedure.

Table \ref{tab:tab4} illustrates the ultimate performance of models trained with two distinct methodologies. Across diverse data distributions, the models demonstrate improvements in key performance metrics including Accuracy, AUC, F1-Score, and TPR. Particularly remarkable is the substantial enhancement in TPR observed within the Non-IID(3) experimental group, escalating from the baseline of 0.9389 to 0.9488. However, it is noteworthy that at this stage, the model's FPR metric trails behind FedAvg, increasing from 0.0438 to 0.0499 under IID data, marking a significant rise.
Fig. \ref{fig:accuracy_comparasion} illustrates the model's performance observed during testing at various checkpoints under the two training methodologies. This depiction allows a nuanced understanding of the impact of enhancement strategies on refining the model's training process. It is inferred that these enhancement strategies effectively bolster the stability of model optimization, thereby alleviating fluctuations in its performance. Notably, in the scenario of Non-IID(2), where training data adheres to a Dirichlet distribution while test data adheres to an independent identical distribution, the optimization impact of enhancement strategies on the model's training process is most pronounced. Simultaneously, it is observed that while these strategies stabilize the model optimization process, they also decelerate the model's convergence during the initial training stages.

The experimental findings elucidate the augmentation in model performance across multiple metrics by initializing client-specific data before each local training iteration. Furthermore, this approach effectively curtails fluctuations in the training process, albeit at the cost of a moderated pace in model convergence.

\begin{figure}[t]
    \centering    \includegraphics[scale=0.36]{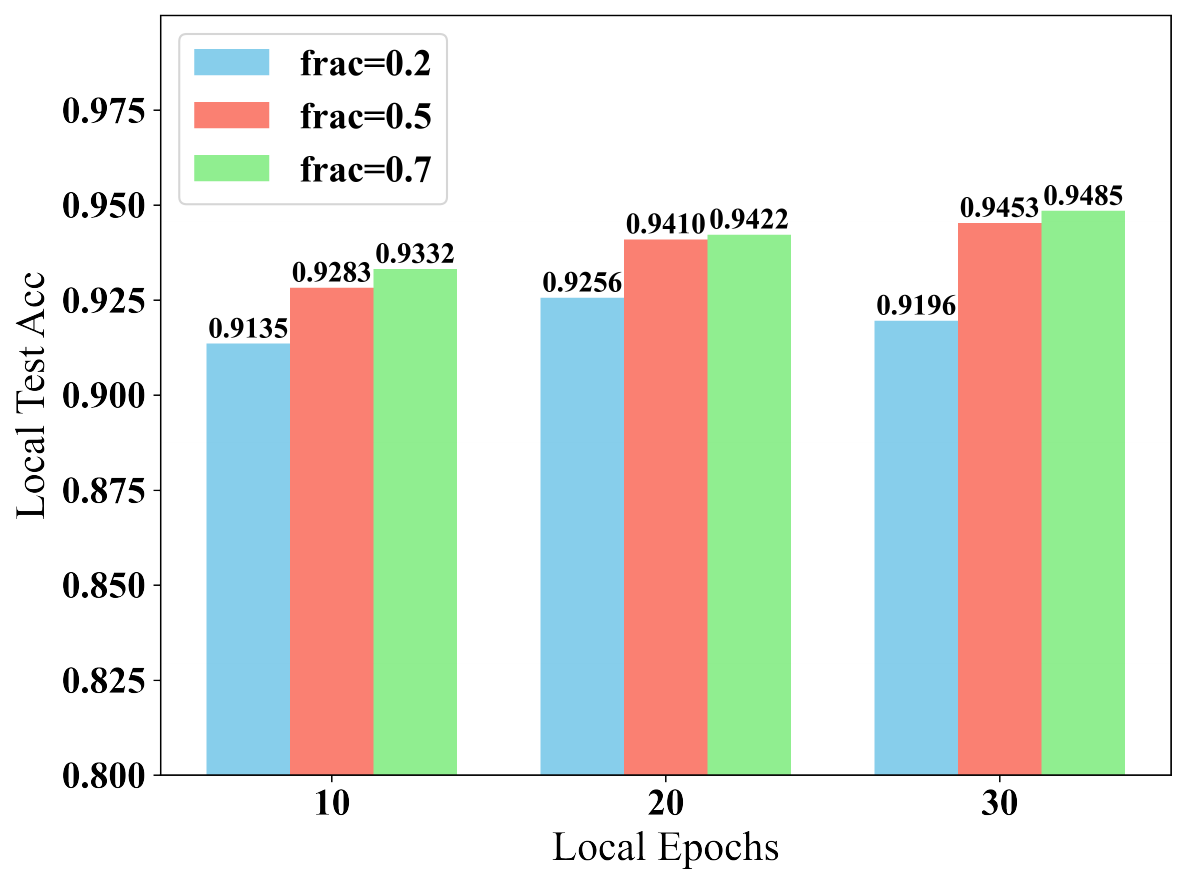}
    \caption{\textbf{Investigation about the Impact of Client Sampling Ratios
on the Model Training Process}}
    \label{fig:frac_ratio}
\end{figure}

\subsubsection{Evaluation of Client Selection Ratio}

Federated learning targets numerous client devices containing training data. Engaging all clients in each epoch ensures higher-quality model training, though, this approach incurs substantial expenses in training costs. Hence, it is imperative to select a suitable proportion wherein a random subset of clients participates in each training epoch, ensuring model training quality while mitigating time and computational overheads. Our work adopts a selection proportion of 0.5 and this experiment aims to validate the effectiveness and rationality of this setting. In our experiments, various experimental groups train using the FedAvg scheme on IID datasets. These groups are trained with proportions of 0.2, 0.5, and 0.7, respectively, and the model's training performance is observed at three distinct time intervals during the training process.

Fig. \ref{fig:frac_ratio} illustrates the model performance in terms of Accuracy after different rounds of training under varying client sampling proportions. It is evident that employing a higher client sampling proportion yields superior model performance after an equivalent number of training rounds. The experimental group with a 0.2 sampling rate notably lags behind the other two groups, exhibiting accuracies of 0.9135, 0.9256, and 0.9195 after 10, 20, and 30 training rounds, respectively. However, the disparities between the experimental groups with 0.5 and 0.7 sampling rates are marginal. At each time point, the Accuracy reaches 0.9283 and 0.9332, 0.9410 and 0.9422, 0.9453 and 0.9485 for 0.5 and 0.7 sampling rates, respectively. Considering the trade-off between the 0.7 sampling rate's time and computational costs alongside the incurred training overheads, a careful balance with the quality of model training suggests that a 0.5 sampling rate presents a more reasonable choice.

\begin{figure}[t]
    \centering
    \includegraphics[scale=0.35]{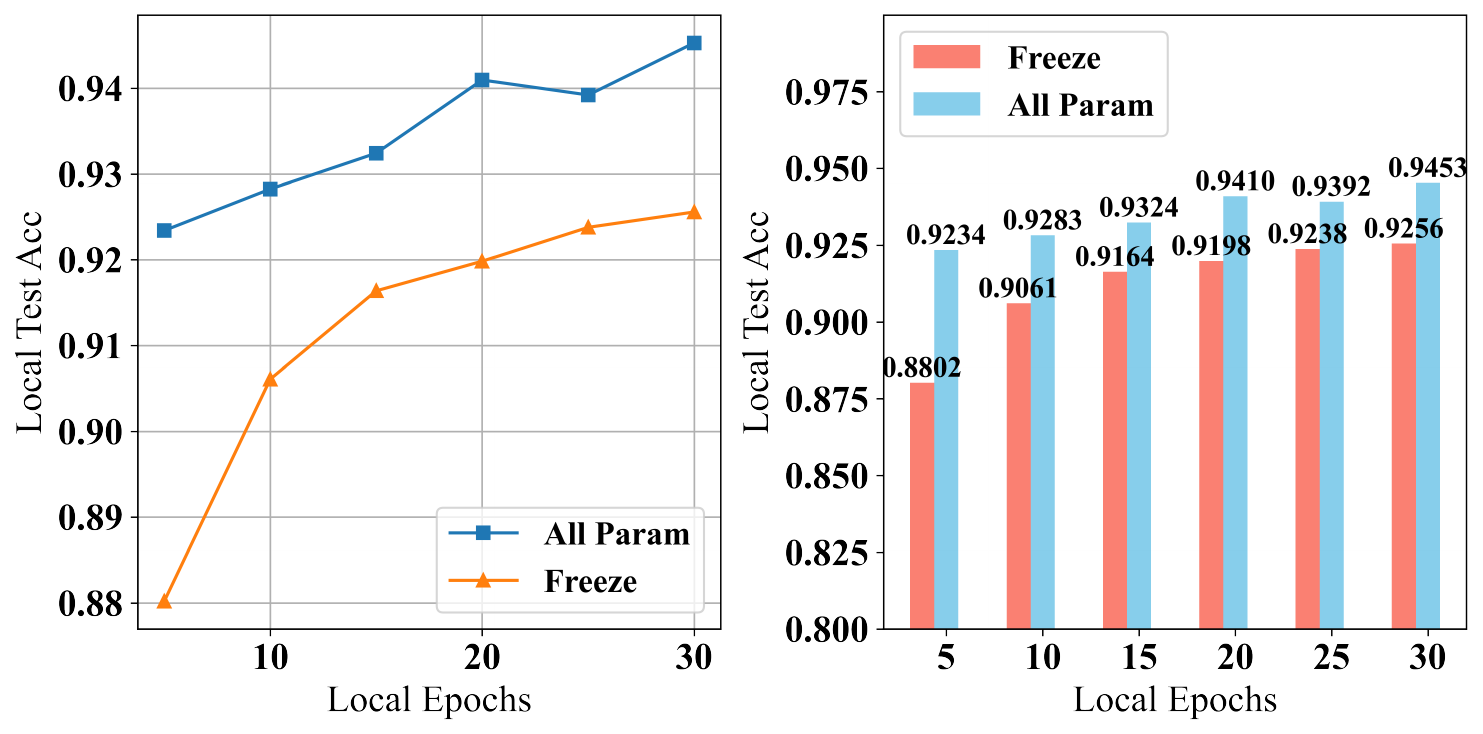}
    \caption{\textbf{Comparison of Model Performance in Fine-Tuning: Local Transformer Layer Freezing vs. Full Parameter Training.  In this comparison, both the training and test sets adhere to an identically distributed setup, and the fine-tuning stage employs the FedAvg algorithm for federated training.(Freeze denotes parameter freezing)}}
    \label{fig:freeze}
\end{figure}

\subsubsection{Study on Parameter Freezing}
During the fine-tuning stage, the FedAvg scheme stipulates training to take place entirely on the clients' local devices while communicating the model's parameter weights with the server. In this scenario, we also account for the heterogeneity present in client devices concerning computational power and memory resources. Considering this, we contemplate the freezing of certain model parameters during training to alleviate the computational and memory overhead on the client side, aiming to enable a larger number of clients to participate in federated training, leveraging the unique data value from heterogeneous client devices and enhancing the richness of the training data. In this experiment, we attempted to freeze certain Transformer layers of the BERT model, significantly reducing the number of parameters involved in training, as compared to a scheme involving all parameters in training. We conducted training according to the FedAvg mechanism on the same IID dataset, comparing and observing the model's training effectiveness at multiple time points. This was done to investigate the performance of the model after freezing parameters.

The experimental results of freezing model parameters are presented in Figure \ref{fig:freeze}, where we observe the model's Accuracy performance across six different time points during the training process. It is evident from the graph that, after the same number of training epochs, the performance of the model with frozen parameters is consistently lower in terms of the Accuracy metric compared to the model trained with all parameters, the gap diminishing as the training progresses. After 30 rounds of federated training, the model with frozen Transformer layers achieved an Accuracy of 0.9256, falling behind the other experimental group though, the performance remained satisfactory. This indicates the effectiveness of employing parameter freezing to enhance client participation in federated training when using Transformer for URL detection tasks. In subsequent research, we believe that adopting a more refined parameter freezing strategy, such as selectively freezing certain self-attention layers within the Transformer, may yield improved training performance.

\section{Discussion}

We have investigated the feasibility of constructing a URL pre-training model in a federated setting and proposed a lightweight client-side federated learning strategy. Our approach has demonstrated comparable performance to the central model in both independent and non-independent identically distributed (Non-IID) scenarios. Here, we provide a brief discussion of our work:
\begin{enumerate}
    \item Performance: Our study primarily investigates whether a federated pre-training Transformer can achieve performance similar to a centrally trained Transformer. With a limited dataset for URL training, significantly smaller than the corpus used for BERT pre-training, we achieved approximately 95\% detection accuracy. Previous studies \cite{10095719,liu2023pyratrans,liu2023malicious} have shown that pre-trained Transformers can attain over 99\% accuracy in malicious URL detection. Therefore, in our federated pre-training architecture, feeding more extensive pre-training data could potentially increase detection accuracy.
    \item Generalizability: For performance comparison, both federated and central models were trained on the same dataset. However, in a real-world application, federated learning architectures, due to their diverse participant base, can access training data that more closely mirrors real-world distribution, encompassing various industries and regions. This diversity enhances the theoretical generalizability and practicality of the federated malicious URL detection models.
    \item ALA Module: In our lightweight client-side federated Transformer, we incorporated an Adaptive Local Aggregation (ALA) module to address client heterogeneity in federated learning. Our experiments showed modest performance improvements with ALA, which is promising, especially considering that our training involved a single, public dataset with minimal client heterogeneity. In real-world scenarios, where client diversity is significantly higher due to industry and regional differences, the ALA strategy has the potential to yield more substantial performance improvements in federated pre-training.
    \item Tokenizer: In our study, for simplicity, we used BERT's tokenizer instead of developing a specialized tokenizer for URLs. Although using BERT's tokenizer is convenient, it is not the most effective approach. Developing a custom tokenizer for URLs is essential for understanding URL semantics and generating domain-aware tokens. Therefore, future research should systematically investigate the development of a specialized tokenizer for URLs.
\end{enumerate}

\section{Conclusion}
In this study, we introduce the first federated pre-trained URL model. Its primary aim is to promote cross-organizational collaboration in the field of malicious link detection by ensuring data privacy and security. This model is designed to meet the strong data demands of large models and allows participants with limited computational resources to benefit from the use of large models while ensuring security. Our experiments demonstrate that the federated pre-trained URL model can achieve results similar to centralized models, indicating its potential in addressing the issue of malicious link detection. From a technical perspective, our federated URL pre-training model achieves lightweight client-side processing, eliminating the need for clients to bear a significant training burden, making it more aligned with practical expectations.

\bibliographystyle{unsrtnat}
\bibliography{reference}
\end{document}